\def\preprint{1}		
\def\comment#1{}
\if\preprint1
	\documentclass[usenatbib]{mn2e}
        \usepackage{natbib}
	\usepackage{times}
	\usepackage{graphicx}
	\usepackage{calc}
\else
	\documentstyle[astrop-bib,referee,times]{mn}
	\newcommand{\includegraphics}[1]{}
\fi

\def\oversim#1#2{\lower0.5pt\vbox{\baselineskip0pt \lineskip-0.5pt
     \ialign{$\mathsurround0pt #1\hfil##\hfil$\crcr#2\crcr\sim\crcr}}}

\title[The  evolution of R~Hya]
{The evolution of the Mira variable R Hydrae}
\author[A.A. Zijlstra et al.]
       {Albert~A.~Zijlstra,$^1$ \thanks{E-mail: \tt a.zijlstra@umist.ac.uk}
       T.~R.~Bedding$^2$\thanks{E-mail: \tt bedding@physics.usyd.edu.au}
        and 
        J.A. Mattei$^3$\thanks{E-mail: \tt jmattei@aavso.org}\\
	$^1$UMIST, Department of Physics, P.O. Box 88, Manchester M60 1QD, UK\\
	$^2$School of Physics, University of Sydney 2006, Australia\\
        $^3$AAVSO, 25 Birch St., Cambridge, MA 02138, U.S.A. \\
}
\pubyear{2002}

\begin{document}

\maketitle

\begin{abstract}

The Mira variable R Hydrae is well known for its declining period,
which Wood \&\ Zarro (1981) attributed to a possible recent thermal
pulse. Here we investigate the long-term period evolution, covering
340 years, going back to its discovery in AD 1662. The data includes
photometric monitoring by amateur and other astronomers over the last
century, and recorded dates of maximum for earlier times. Wavelets are
used to determine both the period and semi-amplitude. We show that the
period decreased linearly between 1770 and 1950; since 1950 the period
has stabilized at 385 days. The semi-amplitude is shown to closely
follow the period evolution. Analysis of the oldest data shows that
before 1770 the period was about 495 days. We find no evidence for an
increasing period during this time as found by Wood \&\ Zarro.  We
discuss the mass-loss history of R~Hya: the IRAS data shows that the
mass loss dropped dramatically around AD 1750. The evolution of the
mass loss as function of period agrees with the mass-loss formalism
from Vassiliadis \&\ Wood; it is much larger than predicted by the
Bl\"ocker law. An outer detached IRAS shell suggests that R~Hya has
experienced mass-loss interruptions before.  The period evolution can
be explained by two models: a thermal pulse occuring around AD 1600,
or an non-linear instability leading to an internal relaxation of the
stellar structure. The elapsed time between the mass-loss decline
giving rise to the outer detached shell, and the recent event, of
approximately 5000\,yr suggests that only one of these events could be
due to a thermal pulse. Further monitoring of R Hya is recommended, as
both models make strong predictions for the future period evolution.
We argue that R~Hya-type events could provide part of the explanation
for the rings seen around some AGB and post-AGB stars.  Changes in
Mira properties were already known on a cycle-to-cycle basis, and on
the thermal-pulse time scale of $\sim 10^4\,\rm yr$.  R Hya shows that
significant evolution can also occur on intermediate time scales of
order $10^2$--$10^3\,\rm yr$.

\end{abstract}

\begin{keywords}
stars: individual: R~Hya
 -- stars: AGB and post-AGB
 -- stars: oscillations 
 -- stars: mass-loss
 -- stars: variables: other 
 -- history and philosophy of astronomy
\end{keywords}

\section{Introduction}

R~Hya is an unusual Mira variable.  Miras are long-period variables
found near the tip of the Asymptotic Giant Branch (AGB).  They show
mono-periodic light curves with large visual amplitudes of more than
2.5 mag. The periods are typically 200--500 days; the amplitude and
shape of the light curve can vary over time but the periods tend to be
stable.  Optical data covering a century or more confirm the
remarkable stability of the Mira pulsations (e.g. \citealt{SBK99}).
But in sharp contrast to this rule, the period of R~Hya has been
declining steadily for over a century\footnote{\citet{Olbers1841}
first noted the irregularity of the period.}.

Although period jitter of a few per cent is common among Miras
\citep{LK93}, possibly related to small changes in the shape of the
light curves, there are only a few examples of significant period
evolution.  Other types of changes appear to be more common, but stars
which exhibit them are automatically classified as semiregular (SR):
the Mira classification requires pulsation stability. The SR class
is a mixture of hidden Miras and non-Mira stars.  Examples of the
former include R~Dor (located on the Mira $PL$ relation) which shows
sudden switches between a period of 330 days and one of 180 days
\citep{BZJF98}, indicative of mode switching. V~Boo has shown an
almost complete disappearance of its Mira pulsation over a century,
albeit without any change in its period \citep{SGK96}. But only R~Aql
is known to show a continuous period decline similar to that of R~Hya.

Early AGB stars contain a helium-burning shell. But during the last
10\%\ of the AGB, when the helium becomes exhausted, the shell
switches to hydrogen burning, punctuated by regular helium flashes:
the thermal pulses \citep{VW93,BS88}. \citet{WZ81} argue that a recent
thermal pulse could explain the period change of R~Hya, if the star is
presently in the luminosity decline following the peak of the pulse.
This interpretation has generally been followed in other papers
discussing period evolution, e.g. on R Cen and T UMi 
\citep*{HMF01,Whi99,MF2000,GS95a}. In support of their interpretation, 
\citet{WZ81} find that the earliest observations  of R~Hya indicate an 
{\em increasing\/} period, which they explain with the luminosity
increase immediately after the onset of the thermal pulse.
 
In this paper we analyse data of R~Hya going back to its discovery in
AD 1662.  The light curve is subjected to wavelet analysis, which
shows how the period and amplitude (the latter available only since
1900) have evolved over time.  We find that the decline in period is
accompanied by a decline in amplitude.  We also find that the period
is no longer decreasing, having stabilised at 385 days in about 1950.
The period was about 495 days before 1770; and we do not confirm the
reported early period increase.  R~Hya appears to have evolved to its
present stable period over approximately 200 yr. 

The paper is organised as follows: In Section~2 we describe the data
and the analysis methods.  Section~3 contains a detailed discussion of
the period evolution.  In Section~4 we discuss the characteristics of
the star. Section 5 discusses the pulsation evolution in terms of
proposed AGB relations. Section~6.1 discusses the mass-loss history
and Section 6.2 describes the two models which can explain R Hya-type
behaviour.  and show that the circumstellar rings observed. Finally,
in Section 6.3 er discuss a possible relation to the rings observed
around AGB and post-AGB stars. The conclusions are summarized in
Section 7.

\section{Observations}

\subsection{The data}

Many bright long-period variable stars have been monitored by amateur
astronomers.  The observations can be found in public archives\footnote{
AFOEV and VSOLJ data are available for immediate download. The BAAVSS data
can be requested by e-mail.  The AAVSO data can be dowloaded for
post-1969 observations and other data can be requested by e-mail. }.  
These archives are valuable resources, and complement high-precision
photometry datasets (e.g., Hipparcos, MACHO) that only cover a few years.

The magnitudes are determined by eye, using reference fields that
contain stars with a range of known magnitudes: a magnitude for the
target star is established by comparison with this standard sequence.
The accuracy is typically 0.1~mag.  For red stars, systematic
differences may exist between observers. When using such data,
sufficient observations from each individual source should be
available to test for individual accuracy and systematic offsets.

\begin{figure*}
\includegraphics[width=\textwidth]{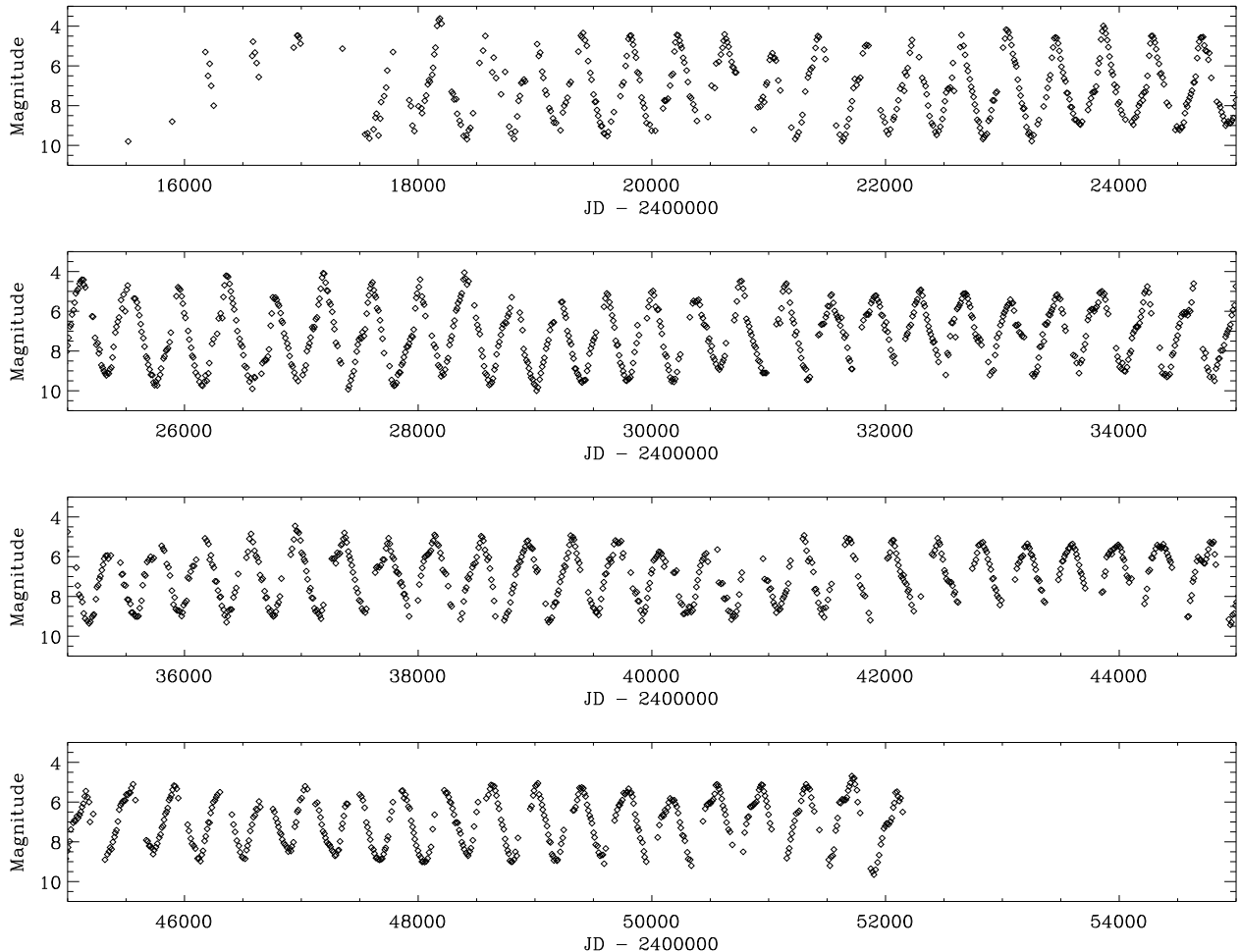}
\caption{\label{rhya.lightcurve} The light curve, using data taken from the
AAVSO (11347 points), AFOEV (1489 points), BAAVS (4369 points) and VSOLJ
(1509 points), binned to 10-day averages}
\end{figure*}

For R~Hya (HR 5080; HIP 65835), we used data from the American
Association of Variable Star Observers (AAVSO), the Variable Star
Observers League of Japan (VSOLJ), the Association Francaise des
Observateurs d'Etoiles Variables (AFOEV) and the British Astronomical
Association, Variable Star Section (BAAVSS)\@.  Only data from
individual observers contributing 30 or more observations were used
and we did not attempt to correct for offsets between observers.
Fig.~\ref{rhya.lightcurve} shows the combined data, binned to 10-day
averages.

Before about AD 1890, the compilations of \citet{CP09} and \citet{MH18}
give derived dates of maxima (and, more rarely, minima), but these do
not include individual measurements.  These data can give a reasonable
estimate for the period if sufficient successive maxima are available.

\subsection{Light curve analysis}

Mira light curves are often analyzed using the so-called O--C
technique (where O stands for the observed date of maximum and C for
the calculated date). The O--C technique must be used with care when
searching for secular period changes because it can be affected by
period jitter: a small jitter can lead to large phase differences over
long time scales \citep{LK93}.  In this paper we prefer the use of
wavelet transforms.

Wavelets are useful when the pulsation properties change over time,
and have been used to study long-period variables (e.g.,
\citealt{SGK96,BZJF98,KSC99}).  We use the weighted wavelet
Z-transform (WWZ: \citealt{Fos96}) developed at AAVSO specifically for 
uneven sampled data.

We experimented with different values for the parameter $c$, which
defines the tradeoff between time resolution and frequency resolution
\citep{Fos96}, and settled on $c=0.005$ as a good compromise.  More
details of the application of the WWZ transform to long-period
variables are given by \citet{BZJF98}.

\section{The period evolution of R~Hya}

\subsection{AD 1850--2001}

Fairly complete coverage is available since 1850, as dates of maxima for
1850--1900 \citep{CP09, MH18}, and as individual magnitude estimates since
1900.  To allow wavelet analysis, the pre-1900 dates of maxima and minima were
arbitrarily assigned magnitudes of 5.0 and 10.0, respectively.
In many cases only dates of maxima were available, and results from the
wavelet analysis could only be obtained by inserting the missing minima.
This was only done when it was clear that consecutive maxima had been
measured, in which case the date of the minimum was taken as being midway
between the maxima.

\begin{figure*}
\includegraphics[clip=true]{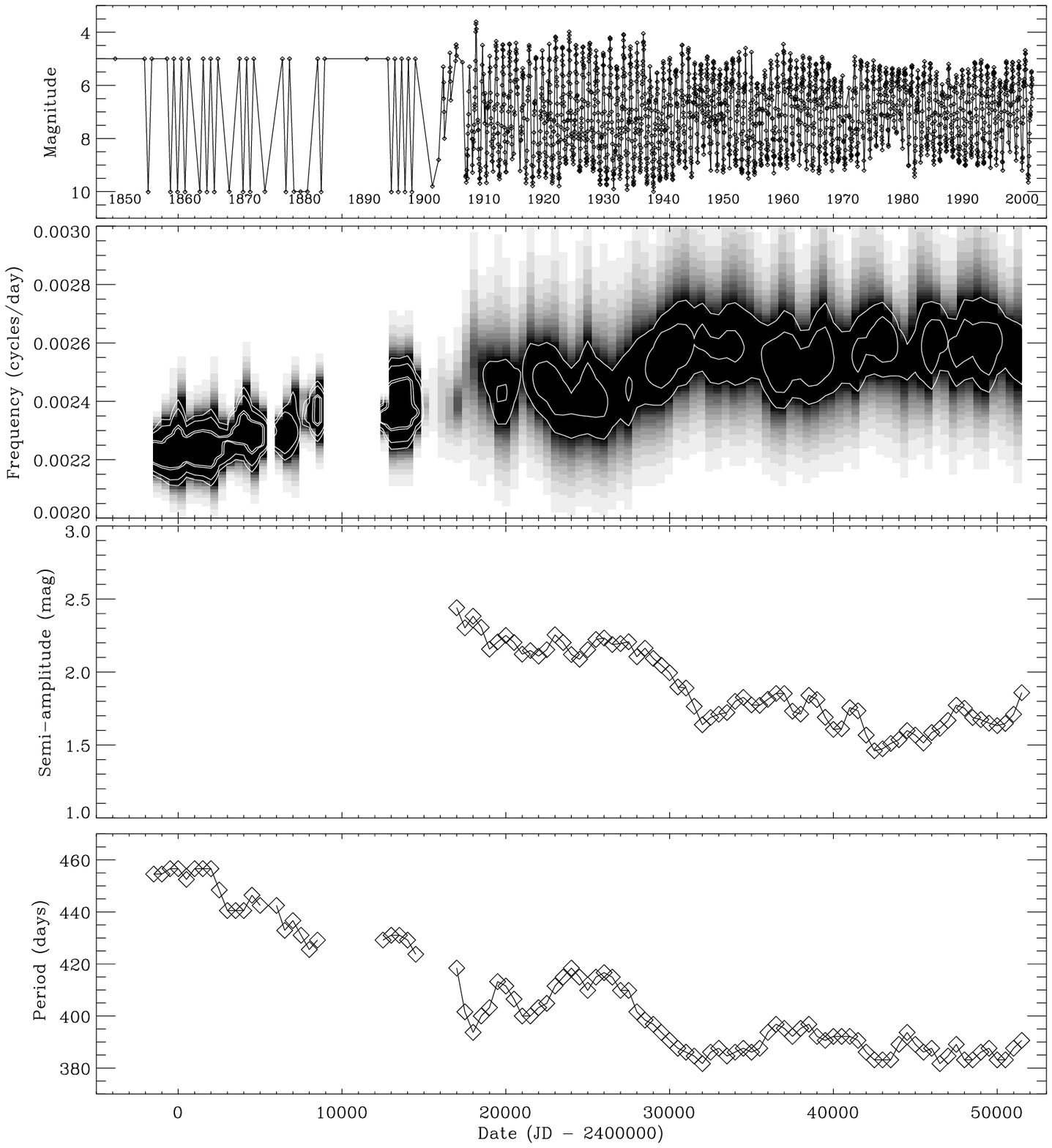}
\caption{\label{rhya.ps} The wavelet analysis for R~Hya, 1850--2000. Shown
are: the light curve, the frequency, the semi-amplitude of the main
frequency component and its period.}
\end{figure*}

Fig.~\ref{rhya.ps} shows the wavelet plot for R~Hya. The lightcurve is
shown in the top panel, and the arbitrary magnitudes assigned to
pre-1900 dates of maxima and minima are obvious.  The second panel
shows the WWZ transform, with the grey scale indicating the
significance of each frequency as a function of time (see
\citealt{BZJF98}).  Only a small range of frequencies is shown --
there was no evidence for significant power outside this range.  The
third and fourth panels show, for each time bin, the semi-amplitude
(in magnitudes) and period (in days) corresponding to the peak of the
WWZ in the second panel.  Note that semi-amplitudes are not available
from these data prior to 1900.

The period evolution in R~Hya is clearly visible, with an overall decline
between 1850 and 1950, from 455 days to 385 days. There is significant
period jitter in addition to the decline.  Before 1900, some of the jitter
may be due to the uncertainty in the dates of maxima. (Different
determinations of the same maximum can typically differ by a few days to a
week, but occasionally much more.)

As made clear in Fig.~\ref{rhya.ps}, the period of R~Hya is no longer
decreasing \citep{Grea2000}.  With hindsight, we can say that the
period stabilized at $\sim$385 days in about 1950, since which time
the period jitter has been limited to the range 380--395 days.  This
jitter is within the normal range of Mira variables \citep{KL95}.  The
period stabilization was preceded by a short phase of rapid decline,
almost 10\% within two decades.

\begin{figure}
\includegraphics[width=9cm]{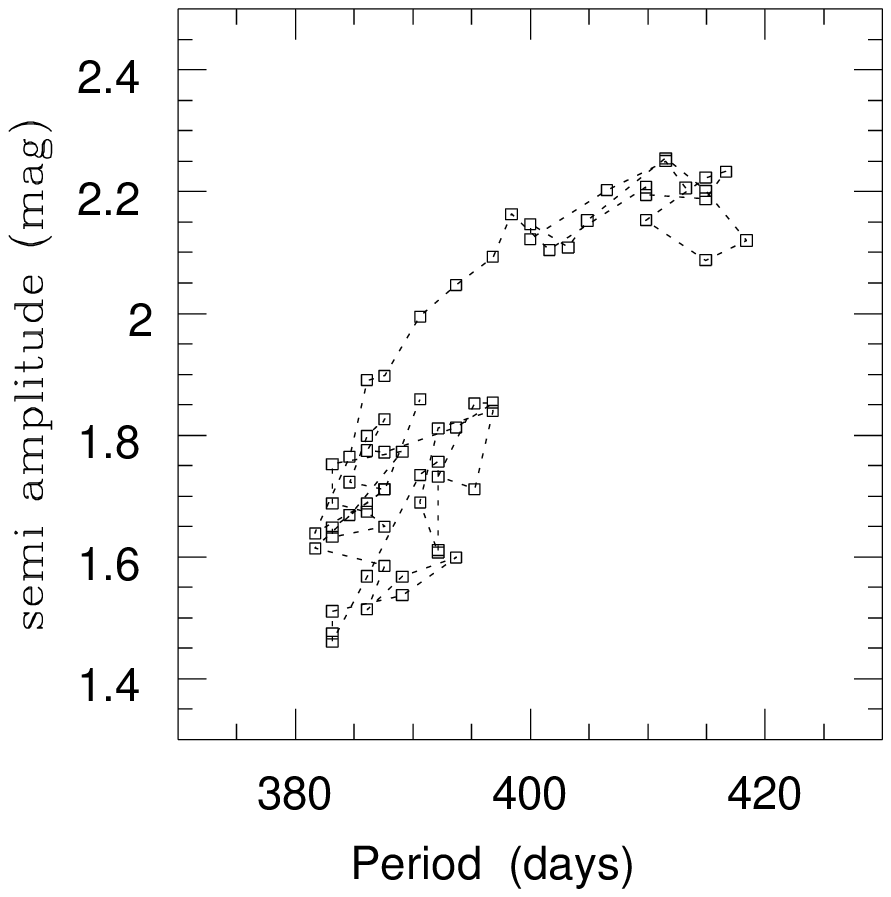}
\caption{\label{rhya.amp} Relation between period and semi-amplitude, from
post-1910 observations. The trend visible in the upper right part of the
plot, of decreasing amplitude with decreasing period, relates to the
1910--1950 data. The group of points in the lower part which show scatter
contain the post-1950 data. }
\end{figure}

Another result is the behaviour in the semi-amplitude.  There was a
decrease from 2.2~mag to 1.7~mag between 1910 and 1950, closely
mimicking the period evolution. The rapid period decline around 1940
is especially well matched by the amplitude, as is the constancy of
the amplitude since 1950. Fig.~\ref{rhya.amp} shows the close relation
between the semi-amplitude and period.  We have reported similar
behaviour in the Mira variables R~Aql, BH~Cru and S~Ori
\citep{BCZ2000} and we speculated that, at least in some cases, the
amplitude changes might {\em cause\/} the period changes via
non-linear effects. For R Hya, the semi-amplitude reached a minimum
around 1975, and has slowly increased again since.

\subsection{AD 1784--1850}

Table~\ref{rhya.old.max} lists all recorded dates of maxima before
1850, based on information given by \citet{MH18}, \citet{A1869} and
\citet{CP09}.  The data are too patchy for wavelet analysis because
the majority of maxima were not observed.  Instead, we determined the
period from observations of maxima that were consecutive or separated
by only a few cycles.  Since the period at 1850 is clearly established
as $\sim$450 days (see above), we start from that date and work
backwards.

\begin{table}
\caption[]{\label{rhya.old.max} Pre-1850 observations of R~Hya. Data
taken from M\"uller \&\ Hartwig (1918) and \citet{CP09} }
\begin{flushleft}
\begin{tabular}{llllllllllll}   \hline
   Year, month,  date of maximum & observer\\
\hline
   1662 04 18 & Hevelius$^a$ \\
   1670/2$^b$ 04 15 & Montanari$^a$ \\
   1704 03 20 & Maraldi$^c$ \\
   1705 09 01: & Maraldi$^d$  \\
   1708 05 20 & Maraldi \\
   1709 11 01: & Maraldi \\
   1712 05 15: & Maraldi \\
   1784 01 26 & Pigott \\
   1785 05 25 & Pigott$^e$ \\
   1805 05 05 & Piazzi \\
   1809 04 04 & Piazzi \\
   1818 03 31: & Olbers $^f$\\
   1823 04 18 & Olbers \\
   1827 01 30 & Schwerd \\
   1843 05 30 & Argelander \\
   1848 04 23 & Argelander$^g$ \\
\hline \\
\end{tabular}
\end{flushleft}
\vskip -10pt
{$^a$ These are dates of observations rather than dates of maxima.
However, given the magnitude range these observations could only have
been made within 1--2 months of maximum.}\\ 
{$^b$ There are two possible dates for this observations (see text).}\\ 
$^c$ Chandler
(Astronomische Nachrichten 2463) derived the five maxima based on
Maraldi's data. Dates in the table are as given by Argelander: Maraldi
gives 1704 March 14 and 1708 May 22.
{$^d$ The uncertain dates of maximum are as given by \citet{MH18}.  The
uncertainties are due to the fact that the maximum was not covered, or
in one case (1712) was difficult to observe because of the Full Moon. We
estimate the uncertainties as 1--2 months. }\\ 
{$^e$ Pigott mentions simultaneous observations by Goodricke which have not 
been published.}\\ 
{$^f$ He also reports observations between 14 March and
4 May 1817, post-maximum, and between 13 Feb and 15 May 1822, also
with an earlier maximum.}\\ 
{$^g$ The date of this maximum is given by Schmidt (independent observation) 
as 3 (or 5) May.}
\end{table}

Argelander's extensive observations gave maxima in 1843 and 1848,
consistent with a period close to 450 days and implying that three
intervening maxima were missed.  Olbers (who discovered the period
evolution) observed maxima in 1818 and 1823, which give a period of
about 460 days (assuming three missing maxima), which also matches the
maximum observed by Schwerd in 1827 (assuming two missing maxima).
Maxima observed by Piazzi in 1805 and 1809 imply a period of 477 days,
assuming two missing maxima.  Finally, the maxima observed by Pigott
in 1784 and 1785 were 485 days apart and were presumably consecutive.

Note that many of the unobserved maxima mentioned above would have
occurred at times of the year when R~Hya was not readily observable
from Europe, as shown in Fig.~\ref{observable.ps}.  Together, these
results suggest that the period of R~Hya was decreasing during
1784--1850 at roughly the same rate as the post-1850 decline, as shown
in Fig.~\ref{rhya.p.full}.

\subsection{AD 1662--1712}

Unfortunately, there is an 80-year gap in observations of R~Hya prior
to those by Pigott in 1784.  The sparse observing record reflects the
poor observability: at the star's southern declination ($-23\deg$),
evening observations from Europe are feasible only in March--May (an
early morning observation is required in winter), and even from Paris
the star never reaches an airmass less than~3.  But the large gap in
the data also coincides with the depths of the Little Ice Age, with
indications for increased cloud cover over Northern Europe \citep{Neu70}.

We are therefore left with the observations of R~Hya made in the first
decades after its discovery, which we now describe in (forward)
chronological order.  The details are taken from papers by \citet{A1869},
\citet{MH18} and \citet{Hoff97}.  The first of these, in particular,
contains a wealth of historical information on several Mira variables.

The first recorded observation of R~Hya was by Johannes Hevelke
(1611--1689; latinized Hevelius), who included it in his second
catalogue \citep{Hevel1679}  but did not note any variability.  The
observations were made from Gdansk, Poland, on the evenings of Tuesday
18 and Wednesday 19 April 1662\footnote{The lutheran Hevelius used the
Gregorian calendar, made clear because he gives the days of the week
of the observations. At this time, the Julian calendar was still in
use in Protestant parts of Europe, but Poland had adopted the
Gregorian calendar in AD 1584, while neighbouring Prussia
had done so in AD 1600. }.

\begin{figure*}
\includegraphics[width=\textwidth]{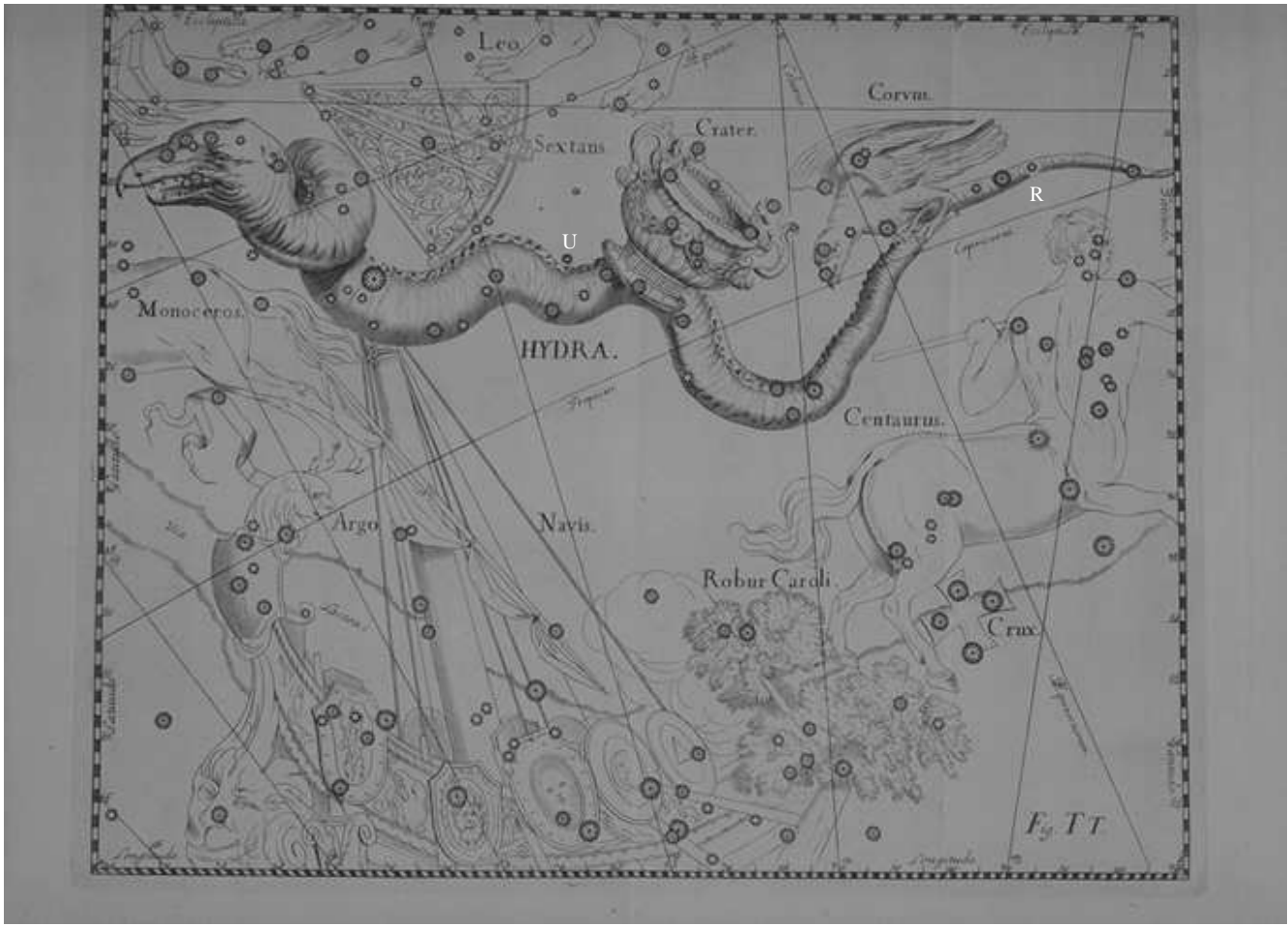}
\caption{\label{Hevel.chart} \citet{Hevel1690a} chart of the region, showing 
both R Hya and U Hya. R Hya is the faint star in the tail of Hydra,
to the right of the tail of the crow: it is just to the right of
$\gamma$ Hydrae and $\psi$ Hydrae. U Hydrae is located between Sextans and
Crater, just above the body of Hydra. }
\end{figure*}

The magnitude found by Hevelius is not certain. In his catalogue
\citep{Hevel1690b}, he
gives it as a 6th mag star, but according to Maraldi (see below)
Hevelius observed it at 5th magnitude \citep{A1869}. Argelander states
that he does not know how Maraldi obtained this value. However,
Maraldi is likely to be correct: R Hya is indicated on Hevelius'
Uranographia \citep{Hevel1690a} (published shortly after this death) as
of similar brightness to $\psi$ Hydra ($m_V=4.97$).  The included
stars are consistent with a brightness limit in this southerly region
at $m_V=5$.  The chart of \citet{Hevel1690a} is reproduced in
Fig. \ref{Hevel.chart}.  In the region below R Hya, a group of stars
with magnitudes of 5.5 and fainter are lacking. \citet{CP09} gave the
discovery observation as the date of maximum, but \citet{A1869} argued
that the real maximum occurred up to 2 months earlier or later. But if
the star was a magnitude brighter than assumed by Argelander, it could
have been closer to maximum\footnote{ Red stars can appear brighter to
the naked eye in conditions of bright moon light, but the discovery
date coincided with new Moon.}.

Given this brightness limit, the variable U Hydrae should also 
be within the range of the catalogue. This carbon star varies between
visual magnitudes of 4.7 and 5.7. It indeed appears to be present in
the chart. There may be other, even older Chinese records of this variable
\citep{Hoff97}, although its semi-regular variability was not discovered 
until 1871. We have not investigated U Hya further, but note that the
stars shown in this region are also consistent with a limit of little
fainter than 5.0.

Geminiano Montanari, independently and before Hevelius' catalogue was
published, observed R~Hya while working at the Paris Observatory.
While comparing Bayer's Uranometry with the sky, he noticed an
unmarked 4$^{\rm th}$ magnitude star along the line connecting $\psi$
and $\gamma$ Hya.  Montanari did not publish the discovery and it is
not known whether he noticed the variability.  (Montanari had
discovered the variability of $\beta$ Per a few years earlier.)  He
entered the star with its magnitude on the map of Bayer.  The date of
his observation was 15 April in either 1670 or 1672 (see below).

\begin{figure}
\includegraphics[height=11cm]{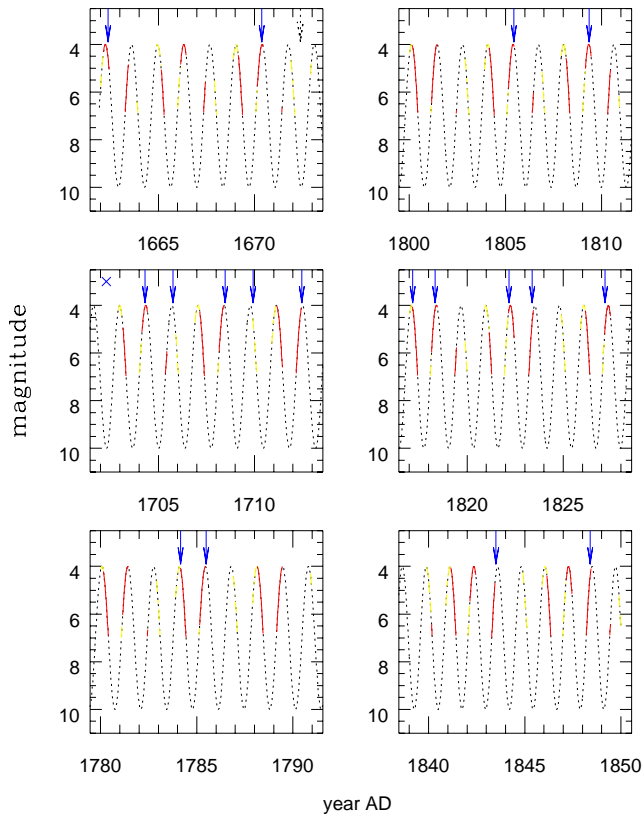}
\caption{\label{observable.ps} The dates of observations are shown with a
sine curve to indicate the proposed variability of R~Hya. The plotted
sine curve has a constant period of 496 days until 1770, declines at
0.4 days per year until 1800 after which the decline is 0.78 days per
year.  (This is not a unique fit, due to the paucity of data.)
The solid curves indicate the parts of the light curve where
observations before midnight would have been possible from Europe: March--June
with $m<7$. The dashed line extends this to include early morning observations.
}
\end{figure}

The variability of R~Hya was first established by Giacomo Filippo
Maraldi (the nephew of Cassini).  In 1702 he tried to re-identify
R~Hya based on Montanari's chart, but failed.  But in March 1704 he
observed the star and followed its appearances and disappearances
until 1712 and identified maxima in 1704 and 1708\footnote{ The later
observations by Maraldi were reported by \citet{Cassini1740}.
}. According to \citet{MH18}, Maraldi suspected a period of two years
but, as they point out, this contradicts his own observations. They
quote Pigott as deriving a period of 494 days in 1786 from his own and
Maraldi's observations.  As we can see in Fig.~\ref{observable.ps},
this period fits the five maxima of 1704--1712 very well.  The period
also agrees with Maraldi's failure to detect the star in 1702, when it
would have been near minimum.

The accuracy of the dates of maximum should not be overstated.  The
high airmass  worsens the effects of the colours of the comparison
stars. \citet{A1869} classifies the maxima according to accuracy, but
for even the best determinations (1784 \&\ 1785, 1823, 1848) he
estimates the uncertainty as 6--7 days.

The evidence seems convincing that the period of R~Hya during the time
of Maraldi was about 495~days.  As can be seen in
Fig.~\ref{rhya.p.full}, this indicates that the rate of period
decrease was much less in the 18th century than in the 19th.

We now turn to the two pre-1700 observations, by Hevelius (in 1662) and
Montanari (in either 1670 or 1672).  The uncertainty in the year of
Montanari's observation is unfortunate.  The observation was published by
Maraldi, along with his own observations, in the Memoires de Paris (pour
l'an 1706 and 1709), where the date was given twice as April 1672 and twice
as 1670.  However, in his own calculations Maraldi consistently used 1670.
On the other hand, Montanari himself did not mention the star in a short
paper of an academic speech from 1671 or 1672, describing several novae,
which could favour the later date\footnote{The paper refers to a book in
preparation on the `Instabilita de firmamento' but this never appeared.}.
\citet{A1869} and \citet{MH18} preferred 1670, while \citet{CP09} gave 1672 as
the more likely.

Assuming that the observation by Hevelius was made close to maximum,
any period close to that from Maraldi's data predicts a minimum in or
around April 1672. On the other hand, accepting 1670 as the correct
date and using a period of 496 days, we find an excellent fit with the
observation by Hevelius and also with those of Maraldi (see
Fig.~\ref{observable.ps}). On their own, the two observations on 1662
and 1670/2 are consistent with an almost unconstrained range of
periods, and they can be made to fit any {\em changing} period. But
the fact that Maraldi's confirmed period also fits these older
observations leads to the plausible hypothesis that the period at the
time of discovery was constant, at about 496 days.  We do not find
support for the statement by \citet{WZ81} that ``four very old
(1662-1708) and valuable dates of maximum ... show that the period was
increasing.''

With this assumption of constant period, all observed maxima from 1662
to 1712 can be fitted to within 3 weeks (with the exception of the
poorly determined maximum in Nov.\ 1709, which is predicted 2 months
earlier).  Maxima would have occurred in February 1662 and in March
1670, in good agreement with actual measurements.  The lack of repeat
observations by Montanari is also explained: the figure shows that the
star would have been difficult to find for 2--3 years after his
observation.

\subsection{The period evolution}

\begin{figure*}
\includegraphics[angle=270,width=\textwidth]{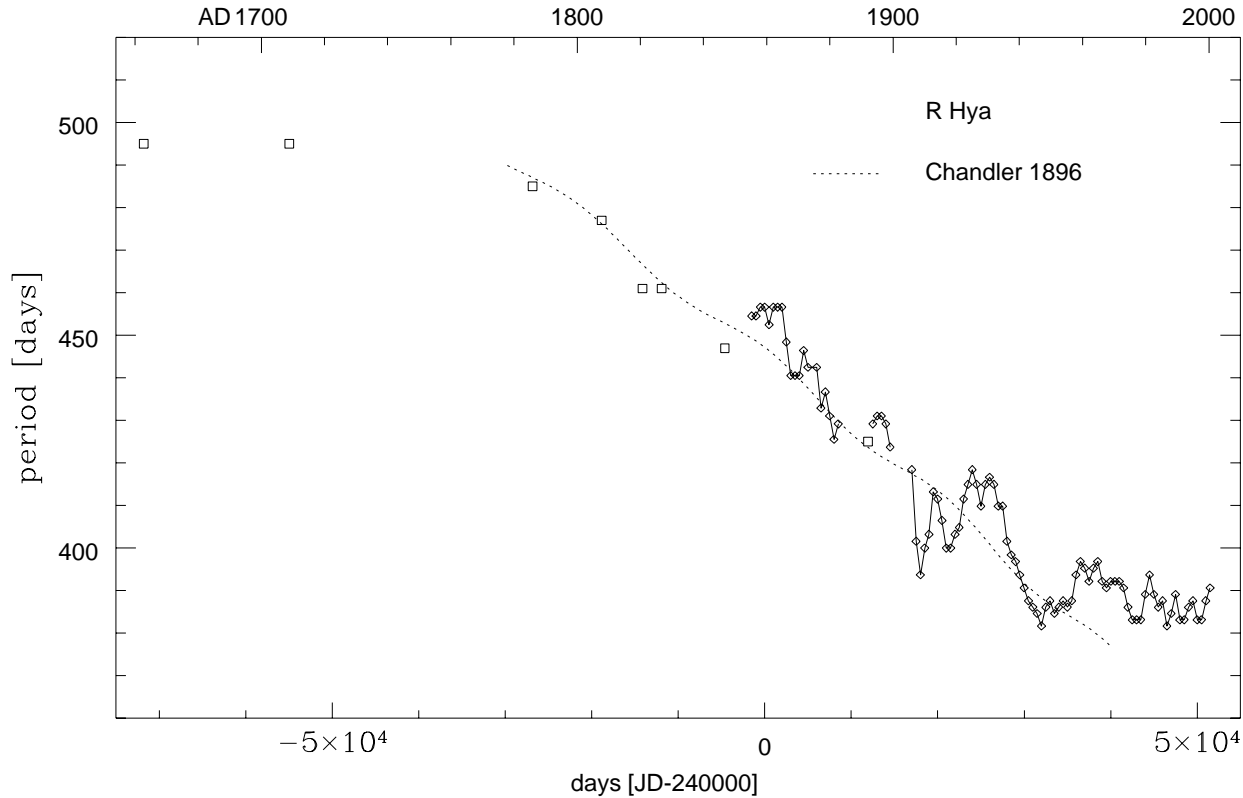}
\caption{\label{rhya.p.full} The period evolution of R~Hya between 1662 and
2001.  The first  point is uncertain; the period is well determined
from 1704 onwards. Extrapolation of the linear decline suggests the decline
began around 1770. The dashed line is the fit proposed by
\citet{Chan1896}.}
\end{figure*}

In summary, the period was approximately 495 days around AD 1700,
declining to 480 days by AD 1800, 450 days by AD 1850, 420 days in AD
1900 and 380 days in AD 1950. The decline was almost linear, at 0.58
days/year: extrapolation suggests that the decline may have started
around 1770, but it is also possible that the decline was initially
slower and began earlier.  Unfortunately, the decline probably began
during the long gap in observations.  The phases of constant period
after and (possibly) before the decline suggest the possibility that
the star has evolved from one (quasi) stable period to another.

Fig.~\ref{observable.ps} shows how the observed dates of maxima fit
with the period evolution.  It is difficult to fit all observations
with a purely linear period decline with a sudden onset. The fit used
in the figure assumes a constant period until 1770, declining by 0.4
days per year between 1770 and 1810, with the decline increasing to
0.6 days per year after 1810. All dates of maxima before 1850 can be
fitted well with this evolution. However, the constraints are
relatively poor and equally good fits are feasible without assuming a
gradual start to the period decline.  Instead the result can be
taken as evidence for some period jitter. The post-maximum
observations of Olbers in 1817 and 1822 are indicated as maxima at 01
Feb of those years.  The figure shows that in both cases, the window
of observability indeed fell post-maximum.

The full period evolution is shown in Fig.~\ref{rhya.p.full}.
The dashed line is the fit proposed by \citet{Chan1896} in the third
catalogue of variable stars. The sinusoidal component is not
confirmed but the slope of his fit gives a good approximation until
the period decline ended around 1950. 

A linear decline implies a constant rate of change in fraction per
cycle, $\delta = 1.6 \times 10^{-3}$.  The time scale of the decline,
defined as $\tau = P /\delta$, where $P$ is the initial (longest)
period, is $\tau = 880\,\rm yr$.  This is an average time scale: the
period evolution also showed significant jitter, with a fastest time
scale (around 1940) of $\tau \approx 200\,\rm yr$.

Amplitude data, available since about 1890, show that the decline in
period was accompanied by a decline in semi-amplitude, from 2.2 to 1.7
mag since 1905, or 5 mmag/cycle. The time scale for the amplitude
decline, extrapolating back to 1770, is about 800 yr, the same as for
the period decline. The relation between period and amplitude is
roughly linear (Fig.~\ref{rhya.amp}).  The visual amplitude of an
oxygen-rich Mira depends on the temperature variation during the
pulsation, leading to the formation of molecules (TiO, VO) during
minimum which strongly absorb at optical wavelengths \citep{RG02}.  A
relation between amplitude and period could therefore be strongly
non-linear, but this is not seen in R~Hya.

\section{Stellar parameters}

The main uncertainty in the mass and luminosity of R~Hya derives from
its uncertain distance.  Unfortunately, the Hipparcos parallax is a
non-detection: $1.6 \pm 2.4$\,mas.  \citet*{WMF00} found a distance of
140\,pc, derived by placing R~Hya on the Mira period--luminosity
($P$--$L$) relation.  \citet{Egg85,Egg66} reported a proper-motion
companion which gave a distance modulus of 6.1 (165\,pc).
\citet{JK92} favoured a distance of 110pc, based on a $P$--$L$
relation.  Only the value of Eggen is consistent with Hipparcos at the
2-$\sigma$ level: these two are also the only direct measurements.

Using $d=165$\,pc, the luminosity is $L=1.16 \times 10^4\,\rm
L_\odot$.  The mean of the Eggen and Hipparcos distances, 400\,pc,
would already yield a luminosity above the classical AGB limit
($M_{\rm bol}=-7.2$: this limit can be exceeded in the case of hot
bottom burning, but only for very large core masses: \citealt{BS91}).
For this reason we will use Eggen's distance in the discussion to
follow.

\citet{Egg85} argued that R~Hya is located within the Hyades supergroup,
with an age of $5$--$10 \times 10^8\,\rm yr$.  This would imply a
progenitor mass for R~Hya around $2\,\rm M_\odot$. The presence of
technetium ($^{99}$Tc) in R~Hya \citep{LLB87} shows that the star is
in the thermal pulsing phase of the AGB (e.g., \citealt{LH99}): this
element is dredged-up during the thermal pulses but has a half life of
$\sim 10^5\,\rm yr$, several times the interpulse time scale. Its
abundance slowly increases during the TP-AGB. Tc is found in 15\%\ of
semiregulars but 75\%\ of long-period Miras.

Infrared photometry was reported by \citet{WMF00}: phase-averaged
magnitudes are $(J,H,K,L)_0 = -1.07, -2.05, -2.45, -2.88)$ and the
bolometric magnitude is $m_{\rm bol}= 0.64$.  The infrared colours
indicate an effective temperature of $T_{\rm eff} = 2830\,\rm K$
\citep{Fea96}.  The $\rm V-K\approx 9.5$  is consistent with
this temperature \citep{BCP98}. \citet{HST95} obtain a lower $T_{\rm
eff} = 2680 \pm 70\,\rm K$ by fitting to the flux distribution between
1.04 and 3.45 microns. They also derive $T_{\rm eff}$ from an angular
diameter measurement: they found $2570\,\rm K$ or $2760\,\rm K$,
assuming fundamental pulsation mode or first overtone, respectively.

The luminosity, with the temperature derived by \citet{Fea96}, yields
a radius of $R \approx 450\,\rm R_\odot$. For our adopted distance,
this predicts an angular diameter of 26\,mas.  The angular diameter
has been measured at 902\,nm as $34.1 \pm 3.4\,\rm mas$
\citep{HST95}, assuming an uniform disk, yielding a  large radius of 
$R=590 \,\rm R_\odot$.  A correction for limb darkening and molecular
opacities brings the value down to about $500 \,\rm R_\odot$ for first
overtone models; for the fundamental mode the effect is much
less. Recently, Tuthill (priv. comm.)  measured a near-infrared
diameter of 24\,mas, in much better agreement with the prediction
above.  Part of the difference between the two observations may be due
to photospheric extensions which can be significant at 902$\,$nm: the
$K$-band is likely to be less affected by this \citep{Fea96}.  In
addition, the earlier observation took place close to minimum
(Tuthill, priv. comm), which in Miras occurs when the star is largest.

\citet{WMF00} and \citet{Fea96} assumed that R~Hya is presently
located on the Mira $P$--$L$ relation. However, this requires a
distance (140\,pc), outside the 2-$\sigma$ confidence limit of
Hipparcos. Given its period history, a location on the narrow $P$--$L$
relation may not be expected. The distance assumed in this paper would
put R~Hya slightly above or to the left of the relation, perhaps
between the Mira and SR branches \citep{BZ98}. The Hipparcos distance
places the star significantly above the relation, a location in
common with O-rich LMC Miras with $P>420\,\rm days$ \citep{FGWC89,
ZLW96}.

\section{The pulsation}

The gradual change in the period of R~Hya implies that its pulsation
mode has remained constant; its evolution is therefore related to a
change in the stellar parameters.  

The pulsation mode of R~Hya is an open question, as it is for all Mira
variables \citep{Woo90, WF00, YT99}. Neither is it proven that R~Hya
exhibits the same pulsation mode as other Miras.  The radius derived
above is consistent with either the fundamental mode or with first
overtone (e.g., \citealt{WF00}: R~Hya falls in between the two modes
in their Fig.~1).

The pulsation equation, which relates the period $P$ (in days) to the radius
$R$ and mass $M$ (in solar units) is given by:
\begin{equation}
 \log P = 1.5\log R -0.5 \log M + \log Q,
\end{equation}
for first overtone pulsators, where the pulsation constant $Q \approx 0.04$ 
\citep{FW82}, 
 or
\begin{equation}
 \log P = 1.949 \log R -0.9 \log M - 2.07
\end{equation}
for fundamental mode pulsators \citep{Woo90}. These equations yield
masses for R~Hya of $0.74\,\rm M_\odot$ and $3.0 \,\rm M_\odot$, respectively.
The large mass required for the fundamental mode provides an argument
for the first overtone, or alternatively for questioning whether the
measured angular diameter is identical to the pulsational diameter.

The two equations both imply that the period evolution was accompanied
by a change in radius: the radius would have decreased by 14--18\%,
depending on pulsation mode. The pre-1770 radius would have been about
$520\,\rm R_\odot$.

There are no direct observations to show how $T_{\rm eff}$ and $L$ 
changed during the period evolution.  \citet{WZ81} fitted a
luminosity decline of 20\%, based on the assumption that R~Hya
underwent a thermal pulse.  \citet{YT96} presented a
different model for period evolution (see below) which does not require
a change in luminosity. The lack of information on the luminosity 
evolution does not allow us to test these two models.

The $P$--$L$ relation derived from LMC Miras is given by:
\begin{equation}
 M_{\rm bol} = -3.00 \log P + \alpha
\end{equation}
\citep{Fea96}. This predicts a luminosity decrease of 25\%
for the period decline of R~Hya.  However, this should be taken as an
upper limit, as R~Hya is unlikely to have evolved along this relation:
the $P$--$L$ relation is not an evolutionary sequence but rather a
sequence of stars with different progenitor masses and metallicities.
The evolutionary tracks of \citet{VW93} cross the $P$--$L$ relation at
almost constant luminosity, while the Whitelock evolutionary track
found in globular clusters \citep{Whi86} is also shallower. But
short-term evolution, such as that shown by R~Hya, may not follow
these sequences either.

Combining the relation between colour and period of
\citet{WMF00},
\begin{equation}
 \left(J - K\right)_0 = -(0.39\pm0.15) + (0.71 \pm 0.06) \log P,
\end{equation}
with the  $T_{\rm eff}$--colour  relation from \citet{Fea96},
\begin{equation}
 \log T_{\rm eff} = -0.59(J - K)_0 + 4.194,
\end{equation}
yields an increase of  $T_{\rm eff}$ for R~Hya of
10\%, i.e., from 2570 to the present 2830 K. Combining this with the
radius change from the pulsation equation gives the counterintuitive
result that the luminosity of R~Hya has increased by 5\%\ rather than
decreased. Given the slope of the $P$--$L$ relation, this suggests that
the slope of the temperature calibration used here is too steep.  The
temperature calibration averages Mira and non-Mira M-type stars.  Using
only Mira variables gives a more shallow relation:
\begin{equation}
 \log T{\rm eff} = -0.474(J - K)_0 + 4.059,
\end{equation}
which gives a 3\%\ decrease in luminosity.  These relations
suggest a negligible change in luminosity. The assumption that R Hya
remained on the AGB colour relations (at constant $L$) may be more
realistic than that of R Hya following $P$--$L$ relations, which
predict decreasing $L$.

\citet{BCP98} give relations between the $V - K$ colour index
and $T_{\rm eff}$ for giants.  The above temperature change implies a
decrease in $V-K$ of about 0.7\,mag.  If R~Hya
evolved along the $K$-band $P$--$L$ relation,
\begin{equation}
  M_K = -3.47 \log P + \beta,
\end{equation}
its $K$-band magnitude would have become fainter by 0.4\,mag.
In this case the $V$-band magnitude should have brightened by 0.3\,mag
since 1770. For the shallower Whitelock track
(e.g., \citealt{BZ98}), the change at $K$ would be less and
the brightening at $V$ closer to 0.7\,mag.

The average visual magnitude has not changed significantly since 1910,
as indicated by the light curve. However, this only covers a fraction
of the period decline. The earliest measurement of Montanari indicated
the star to be of magnitude~4.  R~Hya has not reached this magnitude
during maximum since 1940, but this can be accounted for by the
decline in amplitude and does not imply a change in average
magnitude. It is unlikely that R~Hya was ever much brighter than 4th
mag, because of its absence from the oldest star catalogues.  In
contrast, compare the possible presence of $o$~Ceti in Hipparchus'
catalogue \citep{Cost2002,Manitius1894} (the person, not the
satellite)\footnote{ It is suggested to be the star 'over the
fintails' of Cetus. \citet{MH18} suggest the 'nova' of Hipparchus seen
in 134 BC is $o$~Ceti, but an association with the supernova in Scorpius
\citep{HPY62} appears more likely. }
(but its absence  from the version  in the Almagest \citep{Pto150}),  and
possibly $\chi$~Cygni in Chinese and/or Korean records as a nova on 14
November 1404 \citep{Hoff97}.  But such observations do not allow us
to test the relatively small changes in $V$ predicted above, which in
any case predicts that R~Hya would have been fainter rather than
brighter.

The final assumption we could make is that R~Hya was and remained
on the AGB colour sequence. The AGB equation from \citet{Woo90},
for first overtone pulsation \citep{Fea96}, is given by
\begin{equation}
 M_{\rm bol} = 15.7 \log T_{\rm eff} +1.884 \log z -2.65 \log M -59.1
 -15.7 \Delta
\end{equation}
where the last term represents deviations from the AGB\@.
The relation for fundamental mode is slightly different.  Combining
with the pulsation equation, we find
\begin{equation}
 M_{\rm bol} \propto -2.036 \log P; \qquad 
  \log T_{\rm eff} \propto -0.13 \log P
\end{equation}
\citep{Fea96}. This would yield an increase in $T_{\rm eff}$ 
of 3\%\ and a decrease in $L$ of 20\%. Such changes would be well
within the observational constraints. This parametrized  AGB may
not be valid within the Mira instability strip. Also, if the star is
undergoing a thermal pulse as suggested by \citet{WZ81}, it could be
evolving on a blue loop rather than on the AGB sequence. This would
give a higher temperature and higher (or constant) luminosity. 

It is clear that the various relations are not mutually
consistent.  A luminosity decrease in R~Hya is possible but is not
proven.  Only the radius change, obtained from the period, appears
well constrained.

\section{Discussion}

\subsection{Mass loss evolution: winds of change}

There are strong observational relations between stellar parameters
and mass loss on the TP-AGB\@.  \citet{Blo95} proposed a variation
of the Reimers mass-loss equation:
\begin{equation}
 \dot M_{\rm B} = 4.83 \times 10^{-13} M^{-2.1} L^{2.7} 
 \left(\frac{L R}{M}\right),
\end{equation}
where the last term comes from the Reimers equation \citep{Reim75}.
\citet{VW93} used a very different formulation:
\begin{equation}
 \log \dot M_{\rm VW} = -11.4 +0.0123 P
\end{equation}
for stellar winds below the radiation momentum limit. Both relations
predict a change in $\dot M$ for R~Hya during its recent evolution.
The Bl\"ocker equation predicts, for a change in radius of 15\%\ and
in luminosity of 20\%, that the mass-loss rate would have declined by
a factor of 3. The decline is governed mainly by the luminosity, for
which we have used the most extreme estimate.  If the luminosity has
remained constant, the mass-loss decline would be much smaller.  In
contrast, the \citet{VW93} relation predicts a much steeper decline,
by a factor of 20 independent of any luminosity evolution. Their
relation also predicts a decline of the wind expansion velocity from
14~to 8\,km/s. (Both relations are used to model evolutionary tracks
and may not describe the short-term changes in R~Hya.)

\citet{HIKB98} drew attention to the peculiar IRAS spectrum
of R~Hya, which shows a dust continuum without silicate feature (class
1n). Silicate emission forms close to the star and its lack 
indicates a detached shell.  \citet{HIKB98} derived an
inner radius of $60 R_\ast$, based on a distance of 110\,pc and $R_\ast
= 700 \,\rm R_\odot$. To first order, the inner radius scales with
luminosity.  Scaling to Eggen's distance gives $R_i = 6
\times 10^{15}\,\rm cm$.  For an expansion velocity of 7.5\,km/s 
\citep{WS86}, $R_i$ corresponds to $250\rm \, yr$ BI (before  IRAS).  
This would put the decrease of the mass-loss rate around AD 1750.

The uncertainty in this calculation is significant, not least because
the fit assumes a sudden end to the mass loss, while a gradual
decrease is more likely.  (The mass loss has not ceased completely, as
shown by the presence of an SiO maser \citep{SB75}.)  The outer radius
indicated by the fit is $10^{17}\rm\, cm$, although uncertain.  This
corresponds to an age of 3000\,yr.

The (pre-1770) mass-loss rate derived by \citet{HIKB98}, scaled to
$d=160$\,pc is $\dot M \approx 3 \times 10^{-7}\,\rm M_\odot\,
yr^{-1}$, which is low for a long-period Mira (compare $\dot M_{\rm VW}
= 5 \times 10^{-6}\,\rm M_\odot\, yr^{-1}$). A value around
$10^{-7}\,\rm M_\odot\, yr^{-1}$ is obtained from the CO(2--1)
measurements \citep{WS86}.  \citet{HIKB98} did not give limits on
the present-day mass loss, but the lack of any silicate suggests that
the decline was more than predicted for $\dot M_{\rm B}$, and perhaps
closer to the prediction for $\dot M_{\rm VW}$.

To estimate the required decrease in $\dot M$, we have repeated the
model fit of \citet{HIKB98}. With a single wind, we confirm the
mass-loss rate and cavity size required by the LRS spectrum. If we
fill the cavity with a lower-density wind, a weak silicate feature
re-appears. Only with the new wind at least 10 times less dense can we
fit the spectrum. This is a much larger decrease than predicted by
Bl\"ocker's formalism but is in agreement with the prediction of
\citet{VW93}. The strong decrease predicted by \citet{VW93} appears to
be confirmed for R~Hya.

Interestingly, the IRAS 60-$\mu$m image shows a detached shell around
a bright point source, with an inner radius of 1--2 arcmin (1.5--3$\,
10^{17}\, \rm cm$). \cite{HIKB98} argued that this gap is inconsistent
with their model, with an inner radius that is far too large, and they
cautioned that the deconvolution procedure used (Pyramid Maximum
Entropy) can give artifacts in the presence of a bright point
source. However, the possibility should be considered that 
this ring represents a much older mass-loss event. Its inner
radius indicates that this mass-loss phase was interrupted
$\sim 5000\rm \, yr$ ago.

\subsection{Real-time evolution}

For the time scale on which R~Hya evolves, two models have been
described in the literature that fit its period evolution.

\subsubsection{Post-thermal-pulse evolution}

A thermal pulse occurs when sufficient helium has built up from the
ashes underneath the hydrogen burning layer. The TP gives a strong
modulation of the stellar luminosity.  At first, the luminosity spikes
over a time scale of $\sim 10$--$100,\rm yr$.  Then the luminosity
reaches a short-lived plateau at a level above the hydrogen burning
luminosity (e.g., \citealt{BS88}), followed by a decline on a time
scale of a few hundred years.  The luminosity continues to drop slowly
during quiescent helium burning, reaching around 1/3 of the hydrogen
burning luminosity.  This phase lasts about 10\%\ of the TP cycle.
Finally, helium burning ceases and the hydrogen layer re-ignites,
quickly recovering the pre-pulse luminosity. The period, and also the
mass-loss rate, mimic the luminosity evolution \citep{VW93,
Blo95}. Roughly speaking, the TP phase lasts $10^2$--$10^3$\,yr, the
helium-burning phase $10^3$--$10^4$\,yr and the quiescent
hydrogen-burning phase $10^4$--$10^5$\,yr.  Detached shells around AGB
stars are commonly interpreted in terms of the TP cycle
\citep{ZLWJ92}.

\citet{WZ81} located R~Hya within the earliest post-TP evolution, when
the luminosity shows the steepest drop. In their fit, the peak
luminosity would have occurred around 1750 and the period (and
luminosity) during the Hevelius--Montanari--Maraldi observations would
have been increasing. We have shown that there is no evidence for a
period increase, although it cannot be ruled out either. Sadly, there
are no observations during the crucial phase around 1750. A
near-constant period during 1662--1784 could still be accommodated in
their model by assuming the pulse occurred 50\,yr earlier than assumed
by \citet{WZ81}, placing the peak luminosity plateau around
1700. Their model also predicts a slowing of the luminosity decline
around the present time, which is consistent with the observed lack of
evolution since 1950.

The TP model fits the time scale and period decline well.  A concern
is that it places R~Hya within a unique 100--500\,yr phase of the TP
cycle, corresponding to only 1\%\ of the cycle.  The likelihood of
this occurring in the $3^{\rm rd}$ brightest Mira on the sky is
small. \citet{SCS1937} found continuous period changes in 2 out of 377
well-studied Miras, which is in agreement with this TP-phase.  (A few
more Miras are now known with large period evolution: Bedding et al.,
in preparation).

The duration of the high mass-loss phase pre-1770 may be more
difficult to reconcile with the TP model. If this phase traced the
peak of the pulse, a duration of $\sim 10^2$yr would be expected,
while if it traced the phase of quiescent H-burning it should have
lasted $\sim 10^4$yr or longer. Both the model and the IRAS images of
\citet{HIKB98} suggest it lasted for several $10^3$yr, which is
consistent with neither. 

The evidence for an earlier mass-loss interruption also raises a
problem.  With a time difference of $\sim 5000$\,yr, it is not
possible to relate both to a thermal pulse. If the first event was due
to a thermal pulse \citep{ZLWJ92}, R Hya would presently be nearing
the end of the helium-burning phase or have recently re-entered the
higher-luminosity hydrogen-burning phase, a phase with a much slower
luminospity evolution. For the TP-model, it would be important to
investigate whether the detached ring in the IRAS image is real or
could be explained as an imaging artifact.

\subsubsection{Envelope relaxation}

Mira pulsations are intrinsically non-linear. The period may depend on
the amplitude of the pulsation, affecting either the radius $R$ or the
pulsation constant $Q$. The fact that the amplitude and period of
R~Hya show evidence for simultaneous evolution (see also \citealt{MF2000})
could show the presence
of such a non-linearity. \citet{Woo76} suggests that small variations
in Mira period are most easily explained by an alteration in the
envelope structure near $r = 0.8R_\ast$.

The effect of non-linearity is studied by \citet{YT96}, who calculated
the pulsational stability over a much larger number of cycles than had
been done before. In their models, following an induced perturbation,
the star pulsates in the first overtone for $\sim 200\rm \,
yr$. During this time the growth rate of the fundamental mode is small
but non-zero. Once the fundamental mode begins to dominate, a
re-arrangement of the envelope structure occurs, with entropy
transported downward. The period of the fundamental mode slightly
declines when this mode first dominates, but during the change of the
stellar structure the fundamental period declines over a period of
$\sim 150\rm \, yr$. Their model closest to R~Hya is model D, where
the period first declines to 495 days, and during the restructuring
declines to 330 days. This change is a little larger than seen in R
Hya but occurs on a very similar time scale (but note that the model
star has a much lower luminosity than R~Hya).

The strong points of the model are that the onset, time scale, and
eventual stabilization can all be explained. However, it requires that
the star is initially in a non-equilibrium state and the cause of this
is open. The average luminosity is constant during the period
evolution.

\subsubsection{Cause and effect}

In the TP model, there is a clear cause for the change in period: the
declining luminosity causes a reduction in the stellar radius, which
causes the period to become shorter. In the envelope relaxation model,
what triggers the mode switch is an open question. 

One possibility is the effect of weak chaos. \citet{IFH92} showed that
the outer layers of the star can lose track of the underlying
pulsation and become trapped in `islands of stability'. The effect is
strongest for stars that have reduced envelope masses, and has been
invoked to explain the mode switching in R Dor \citep{BZJF98}.  In the
model of \citeauthor{IFH92}, there is an underlying piston moving with
constant frequency. In real Miras, the non-linearity discussed by
\citet{YT96} implies that if a star is caught in an island of enlarged
radius, over time the inner structure of the star could be affected by
this.  This could act as the trigger for the mode evolution.

Interaction between the star and its extended atmosphere may also have
some effect: \citet{HD97} have shown that feedback from atmosphere on
the star can affect the cycle-to-cycle amplitudes.

We find a clear relation between amplitude and period for R Hya.
\citet{BCZ2000} have suggested that the change in ampitude may
act as the {\it cause} of the period change.

\subsection{Rings}

Several post-AGB stars and one AGB star show concentric rings seen in
reflected light \citep{KSS01}. The separation of the rings (or arcs:
only the illuminated parts are seen) correspond to time scales of
about 500\,yr. The thicknesses of the rings correspond to 0.1--0.5 of
the separation, and the density enhancement in the rings is at least
30\%, but could be larger.  The obvious explanation of these rings is
that the mass-loss rate showed a fluctuation on this time scale
\citep{Sahai98}. However, the only effect known to modify the AGB star
properties on this time scale is the thermal pulse, and this could
only lead to a single ring.

The time scale for the period decline in R~Hya is remarkably similar
to the time scale of the rings. The strong decline in mass-loss rate
following the onset of the period decline makes it the only {\it
observed} Mira behaviour which can explain the rings. However, this
requires the evolution discussed in this paper to be periodic. The
fact that the period has now stabilized allows for the possibility that
it will at some time increase again, but there is at present no direct
evidence that the period evolution is periodic.

\begin{figure}
\includegraphics[width=8.5cm]{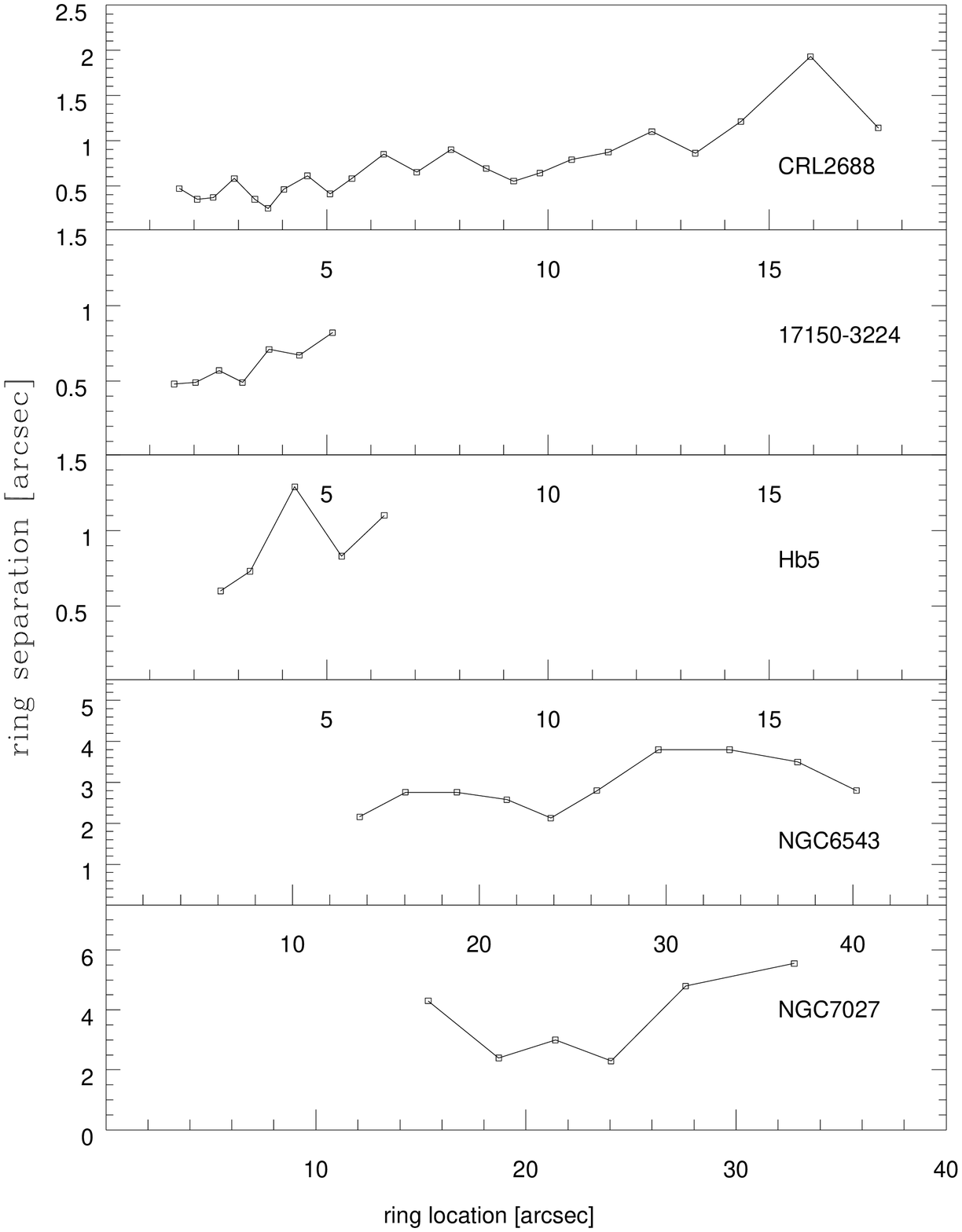}
\caption{\label{rings.ps} Relation between the separation of
the rings observed in post-AGB stars, and the location of the rings,
both in arcsec. The squares corresponds to poitions midway between the
density enhancements.}
\end{figure}

Of the models for the R~Hya evolution discussed above, only the
relaxation model combined with a periodic or stochastic trigger could
lead to the formation of multiple rings.  The observed separation in
the rings is not fully regular: this is shown in Fig.~\ref{rings.ps},
using data taken from \citet{KSS01}.  The separation can vary by as
much as a factor of 2 (although in a few cases an intervening ring may
have been missed). There is also a clear indication for an increase of
rings separation with distance from the star. This implies that the
event causing the rings occured at decreasing time intervals as the
star approached the end of the AGB.

The chaotic behaviour predicted by \citet{IFH92} increases as the
envelope mass reduces, and this behaviour fits both the irregularity
and the increasing frequency of the mass-loss epsiodes. But its effect
on the mass loss is not clear, and it is not proven (although
possible) that this chaotic behaviour can act as a trigger for an R
Hya-type event.

The TP model makes a very clear prediction for the future period
evolution. Further monitoring of R Hya is therefore important:
a continuing but slow decline would agree with the TP model. If, in
contrast, the period is found to increase again, this would rule out
the TP model and make a connection with the post-AGB rings more likely.

\section{Conclusions}

We have studied the period evolution of R Hya, using both magnitude
estimates for the light curve and old data giving dates of
minimum/maximum.  The wavelets are shown to be a powerful tool to
analyse such datasets.  The main results are

\begin{enumerate}

\item The period of R Hya has declined continuously from 495 days to
385 days, between approximately AD 1770 and AD 1950. Before 1770
there is no evidence for period evolution, while after 1950 the period
has been stable, showing at most minor period jitter. The
evolution gives the impression of a change between two relatively stable
configurations. We do not confirm the suggestion that prior to 1770
the period was increasing.

\item The amplitute (available after 1900) closely followed the period 
evolution, declining at first but becoming stable after 1950. A
relation between amplitude and period is typical for a non-linear
pulsation.

\item The most likely distance is 165 pc, giving a luminosity of $
1.16 \times 10^4\,\rm L_\odot$.  The likely progenitor mass is around
$2 \,\rm M_\odot$. The star is located on the thermal-pulsing tip of
the AGB.

\item  The period change indicates a decrease in stellar radius. The 
luminosity and temperature change is less secure. Assuming the star
remained on the fiducial AGB relations, the temperature change may
have been 10--20\%.  Various luminosity-dependent AGB relations
predict changes in the luminosity ranging from 25\%\ decrease to 3\%\
increase. Given the uncertainty whether R Hya satisfied such relations
during its period decline, and the fact that different relations do
not even agree on the sign, it is not possible to confirm that the
luminosity decreased: a constant luminosity is a significant
possibility.

\item The IRAS spectrum shows that mass-loss rate has recently declined by 
a factor of at least 10. A model of the IRAS spectrum shows that the
mass-loss decline occured about 250 yr ago. This is in good agreement
with the onset of the period decrease and suggests the two effects are
correlated.  The pre-1770 mass-loss rate was $\dot M \approx 3 \times
10^{-7}\,\rm M_\odot\, yr^{-1}$. A large detached IRAS shell suggests
an earlier phase of high mass loss, ending about 5000 yr ago. The
post-1770 decline agrees with the $\dot M$--$P$ relation of
Vassiliadis \&\ Wood (1992) but is much larger than predicted from the
mass-loss formalism of Bl\"ocker (1995).

\item Two models can explain the behaviour of R Hya. First, a recent thermal 
pulse, occuring shortly before the discovery. This also can fit the
constant period since 1950.  The second model is envelope relaxation,
where the non-linearity of the Mira pulsation causes a change in the
entropy structure of the star.  Both the period evolution between two
semi-stable states and the time scale of the change are
reproduced. There is at present insufficient data to decide between
the two models.

\item The evidence for a strong effect on the mass loss raises the
possiblity of a connection with the circumstellar rings observed
around some post-AGB stars. The evolution seen in R Hya is the only
observed effect in Miras which has the correct time scale.  Icke et
al. (1992) show that Mira period instability increases as the envelope
mass decreases.  This would place such events at the tip of the AGB,
and would agree with the observations that the time scales between
'ring' events decreases with time. However, a mechanism to translate this
chaotic envelope behaviour into structural (period) evolution of the
Mira is lacking.

\item Further monitoring of R Hya is recommended. The thermal-pulse
model makes a strong prediction for its future period evolution. If,
on the other hand, the period at some time would increase again, this
would rule out this model and also make a connection with the post-AGB
rings more likely.

\end{enumerate}

Changes in Mira properties were already known on a cycle-to-cycle basis,
and on time scales of $10^4\,\rm yr$, which is the thermal-pulse time scale.
R Hya shows that significant evolution can also occur on intermediate time
scales of order $10^2$--$10^3\,\rm yr$.

\section*{Acknowledgments}

We thank the many amateur observers and those who maintain the
databases of the AAVSO, AFOEV, BAAVSS and VSOLJ\@.  The large amount
of effort of the many amateur astronomers has resulted in a highly
valuable and unique tool for studying stellar evolution.  Krzysztof
Gesicki found (and translated) the original Hevelius catalogue.  TRB
is grateful to the Australian Research Council for financial support.
This research was supported by PPARC, and by an ESO visitor grant. We
would especially like to thank the ESO librarians and Reinhard E.
Schielicke for their valuable help in locating and prvoding access to
the historical records for R Hya.

\bibliographystyle{mn2e}

\bibliography{MB1075}

\end{document}